\documentclass[aps,prl,twocolumn,showpacs,groupedaddress]{revtex4}  % for review and submission 
\usepackage{graphicx}  % needed for figures
\usepackage{dcolumn}   % needed for some tables
\usepackage{bm}        % for math
\usepackage{amssymb}   % for math

\begin{document}

% the following line is for submission 
\hspace{5.2in} \mbox{Fermilab-Pub-05/217-E}

\newcommand{\dzero}     {D\O}
\newcommand{\ttbar}     {\mbox{$t\bar{t}$}}
\newcommand{\qqbar}     {\mbox{$q\bar{q}$}}
\newcommand{\ppbar}     {\mbox{$p\bar{p}$}}
\newcommand{\met}       {\mbox{$\not\!\!E_T$}}
\newcommand{\rar}       {\rightarrow}
\newcommand{\mtop}	{\mbox{$m_t$}}

\title{Measurement of the \ttbar\ Production Cross Section in \ppbar\ Collisions at $\sqrt{s}=1.96$ TeV in Dilepton Final States}
% LIST_OF_AUTHORS_R2.TEX                 5/17/05            
%
\author{                                                                      
%% names begin here                                                           
V.M.~Abazov,$^{35}$                                                           
B.~Abbott,$^{72}$                                                             
M.~Abolins,$^{63}$                                                            
B.S.~Acharya,$^{29}$                                                          
M.~Adams,$^{50}$                                                              
T.~Adams,$^{48}$                                                              
M.~Agelou,$^{18}$                                                             
J.-L.~Agram,$^{19}$                                                           
S.H.~Ahn,$^{31}$                                                              
M.~Ahsan,$^{57}$
% ...
\\                                                                            
\vskip 0.30cm                                                                 
\centerline{(D\O\ Collaboration)}                                             
\vskip 0.30cm                                                                 
}                                                                             
\affiliation{                                                                 
\centerline{$^{1}$Universidad de Buenos Aires, Buenos Aires, Argentina}       
\centerline{$^{2}$LAFEX, Centro Brasileiro de Pesquisas F{\'\i}sicas,         
                  Rio de Janeiro, Brazil}                                     
\centerline{$^{3}$Universidade do Estado do Rio de Janeiro,                   
                  Rio de Janeiro, Brazil}                                     
\centerline{$^{4}$Instituto de F\'{\i}sica Te\'orica, Universidade            
                  Estadual Paulista, S\~ao Paulo, Brazil}                     
\centerline{$^{5}$University of Alberta, Edmonton, Alberta, Canada,           
               Simon Fraser University, Burnaby, British Columbia, Canada,}   
\centerline{York University, Toronto, Ontario, Canada, and                    
         McGill University, Montreal, Quebec, Canada}                         
\centerline{$^{6}$Institute of High Energy Physics, Beijing,                  
                  People's Republic of China}                                 
\centerline{$^{7}$University of Science and Technology of China, Hefei,       
                  People's Republic of China}                                 
\centerline{$^{8}$Universidad de los Andes, Bogot\'{a}, Colombia}             
\centerline{$^{9}$Center for Particle Physics, Charles University,            
                  Prague, Czech Republic}                                     
\centerline{$^{10}$Czech Technical University, Prague, Czech Republic}        
\centerline{$^{11}$Center for Particle Physics, Institute of Physics,         
                   Academy of Sciences of the Czech Republic,                 
                   Prague, Czech Republic}                                    
\centerline{$^{12}$Universidad San Francisco de Quito, Quito, Ecuador}        
\centerline{$^{13}$Laboratoire de Physique Corpusculaire, IN2P3-CNRS,         
                  Universit\'e Blaise Pascal, Clermont-Ferrand, France}       
\centerline{$^{14}$Laboratoire de Physique Subatomique et de Cosmologie,      
                  IN2P3-CNRS, Universite de Grenoble 1, Grenoble, France}     
\centerline{$^{15}$CPPM, IN2P3-CNRS, Universit\'e de la M\'editerran\'ee,     
                  Marseille, France}                                          
\centerline{$^{16}$IN2P3-CNRS, Laboratoire de l'Acc\'el\'erateur              
                  Lin\'eaire, Orsay, France}                                  
\centerline{$^{17}$LPNHE, IN2P3-CNRS, Universit\'es Paris VI and VII,         
                  Paris, France}                                              
\centerline{$^{18}$DAPNIA/Service de Physique des Particules, CEA, Saclay,    
                  France}                                                     
\centerline{$^{19}$IReS, IN2P3-CNRS, Universit\'e Louis Pasteur, Strasbourg,  
                France, and Universit\'e de Haute Alsace, Mulhouse, France}   
\centerline{$^{20}$Institut de Physique Nucl\'eaire de Lyon, IN2P3-CNRS,      
                   Universit\'e Claude Bernard, Villeurbanne, France}         
\centerline{$^{21}$III. Physikalisches Institut A, RWTH Aachen,               
                   Aachen, Germany}                                           
\centerline{$^{22}$Physikalisches Institut, Universit{\"a}t Bonn,             
                  Bonn, Germany}                                              
\centerline{$^{23}$Physikalisches Institut, Universit{\"a}t Freiburg,         
                  Freiburg, Germany}                                          
\centerline{$^{24}$Institut f{\"u}r Physik, Universit{\"a}t Mainz,            
                  Mainz, Germany}                                             
\centerline{$^{25}$Ludwig-Maximilians-Universit{\"a}t M{\"u}nchen,            
                   M{\"u}nchen, Germany}                                      
\centerline{$^{26}$Fachbereich Physik, University of Wuppertal,               
                   Wuppertal, Germany}                                        
\centerline{$^{27}$Panjab University, Chandigarh, India}                      
\centerline{$^{28}$Delhi University, Delhi, India}                            
\centerline{$^{29}$Tata Institute of Fundamental Research, Mumbai, India}     
\centerline{$^{30}$University College Dublin, Dublin, Ireland}                
\centerline{$^{31}$Korea Detector Laboratory, Korea University,               
                   Seoul, Korea}                                              
\centerline{$^{32}$CINVESTAV, Mexico City, Mexico}                            
\centerline{$^{33}$FOM-Institute NIKHEF and University of                     
                  Amsterdam/NIKHEF, Amsterdam, The Netherlands}               
\centerline{$^{34}$Radboud University Nijmegen/NIKHEF, Nijmegen, The          
                  Netherlands}                                                
\centerline{$^{35}$Joint Institute for Nuclear Research, Dubna, Russia}       
\centerline{$^{36}$Institute for Theoretical and Experimental Physics,        
                  Moscow, Russia}                                             
\centerline{$^{37}$Moscow State University, Moscow, Russia}                   
\centerline{$^{38}$Institute for High Energy Physics, Protvino, Russia}       
\centerline{$^{39}$Petersburg Nuclear Physics Institute,                      
                   St. Petersburg, Russia}                                    
\centerline{$^{40}$Lund University, Lund, Sweden, Royal Institute of          
                   Technology and Stockholm University, Stockholm,            
                   Sweden, and}                                               
\centerline{Uppsala University, Uppsala, Sweden}                              
\centerline{$^{41}$Lancaster University, Lancaster, United Kingdom}           
\centerline{$^{42}$Imperial College, London, United Kingdom}                  
\centerline{$^{43}$University of Manchester, Manchester, United Kingdom}      
\centerline{$^{44}$University of Arizona, Tucson, Arizona 85721, USA}         
\centerline{$^{45}$Lawrence Berkeley National Laboratory and University of    
                  California, Berkeley, California 94720, USA}                
\centerline{$^{46}$California State University, Fresno, California 93740, USA}
\centerline{$^{47}$University of California, Riverside, California 92521, USA}
\centerline{$^{48}$Florida State University, Tallahassee, Florida 32306, USA} 
\centerline{$^{49}$Fermi National Accelerator Laboratory, Batavia,            
                   Illinois 60510, USA}                                       
\centerline{$^{50}$University of Illinois at Chicago, Chicago,                
                   Illinois 60607, USA}                                       
\centerline{$^{51}$Northern Illinois University, DeKalb, Illinois 60115, USA} 
\centerline{$^{52}$Northwestern University, Evanston, Illinois 60208, USA}    
\centerline{$^{53}$Indiana University, Bloomington, Indiana 47405, USA}       
\centerline{$^{54}$University of Notre Dame, Notre Dame, Indiana 46556, USA}  
\centerline{$^{55}$Iowa State University, Ames, Iowa 50011, USA}              
\centerline{$^{56}$University of Kansas, Lawrence, Kansas 66045, USA}         
\centerline{$^{57}$Kansas State University, Manhattan, Kansas 66506, USA}     
\centerline{$^{58}$Louisiana Tech University, Ruston, Louisiana 71272, USA}   
\centerline{$^{59}$University of Maryland, College Park, Maryland 20742, USA} 
\centerline{$^{60}$Boston University, Boston, Massachusetts 02215, USA}       
\centerline{$^{61}$Northeastern University, Boston, Massachusetts 02115, USA} 
\centerline{$^{62}$University of Michigan, Ann Arbor, Michigan 48109, USA}    
\centerline{$^{63}$Michigan State University, East Lansing, Michigan 48824,   
                   USA}                                                       
\centerline{$^{64}$University of Mississippi, University, Mississippi 38677,  
                   USA}                                                       
\centerline{$^{65}$University of Nebraska, Lincoln, Nebraska 68588, USA}      
\centerline{$^{66}$Princeton University, Princeton, New Jersey 08544, USA}    
\centerline{$^{67}$Columbia University, New York, New York 10027, USA}        
\centerline{$^{68}$University of Rochester, Rochester, New York 14627, USA}   
\centerline{$^{69}$State University of New York, Stony Brook,                 
                   New York 11794, USA}                                       
\centerline{$^{70}$Brookhaven National Laboratory, Upton, New York 11973, USA}
\centerline{$^{71}$Langston University, Langston, Oklahoma 73050, USA}        
\centerline{$^{72}$University of Oklahoma, Norman, Oklahoma 73019, USA}       
\centerline{$^{73}$Brown University, Providence, Rhode Island 02912, USA}     
\centerline{$^{74}$University of Texas, Arlington, Texas 76019, USA}          
\centerline{$^{75}$Southern Methodist University, Dallas, Texas 75275, USA}   
\centerline{$^{76}$Rice University, Houston, Texas 77005, USA}                
\centerline{$^{77}$University of Virginia, Charlottesville, Virginia 22901,   
                   USA}                                                       
\centerline{$^{78}$University of Washington, Seattle, Washington 98195, USA}  
}                                                                             
%end                                                                          
  % input Dzero author list
%\date{\today}
\date{May 26, 2005}

\begin{abstract}
We present a measurement of the top quark pair ($\ttbar$) production cross section
in \ppbar\ collisions at $\sqrt{s} = 1.96$ TeV using events with two charged leptons
in the final state.  This analysis utilizes an integrated luminosity of 224-243 pb$^{-1}$ 
collected with the \dzero\ 
detector at the Fermilab Tevatron Collider. We observe 13 events in
the $e^+e^-$, $e\mu$ and $\mu^+\mu^-$ channels with an expected
background of $3.2 \pm 0.7$ events. For a top quark mass of 175 GeV, we measure
a \ttbar ~production cross section of $\sigma_{t\bar{t}} =
8.6_{-2.7}^{+3.2}(\rm stat)\pm1.1(\rm syst)\pm0.6(lumi)$ pb, consistent 
with the standard model prediction.
\end{abstract}

\maketitle

The top quark was discovered \cite{disc} in 1995 at the Fermilab Tevatron 
Collider in \ppbar\ collisions at $\sqrt{s} = 1.8$~TeV. Its observation 
completed the third quark weak isospin doublet suggested by the absence 
of flavor changing neutral current interactions \cite{FCNC} and
measurement of the $b$ quark weak isospin \cite{T3b}. By virtue of its 
large mass ($m_t=178.0 \pm 4.3$ GeV \cite{mtopcombo}), the top quark could
decay into exotic particles, e.g. a charged Higgs boson~\cite{charged_higgs}. 
Such decays would lead to a measured \ttbar\ production cross section 
($\sigma_{t\bar{t}}$) apparently dependent on the \ttbar\ final state. 
It is therefore 
necessary to precisely measure $\sigma_{t\bar{t}}$ in 
all decay channels and compare it with the standard model
prediction. The increased luminosity and higher collision 
energy of $\sqrt{s}=1.96$ TeV at the Run II of Tevatron permit substantially 
more accurate measurement of $\sigma_{t\bar{t}}$ in all final states.

%%%%%%%%%%%%%%%%%%%%%%

In the $SU(2)\times U(1)$ electroweak model with one Higgs doublet \cite{S-W},
each top quark of a $t \bar t$ pair is expected to decay approximately 
99.8\% of the time to a $W$ boson and a $b$ quark~\cite{PDG}. Dilepton final 
states arise when both $W$ bosons decay leptonically. These occur along with 
two energetic jets resulting from the hadronization of the $b$ quarks and 
missing transverse energy (\met) from the high transverse momentum ($p_T$) 
neutrinos. In this Letter, we present a measurement of 
$\sigma_{t\bar{t}}$ with 224-243~pb$^{-1}$ of \ppbar\ collider data at 
$\sqrt{s}=1.96$ TeV collected with the upgraded \dzero\ detector \cite{run2det}.
We consider the $e^{+}e^{-}$, 
$e\mu$ and $\mu^{+}\mu^{-}$ final states.
The electrons and muons may originate 
either directly from a $W$ boson or indirectly from a 
$W\rightarrow\tau \nu$ ~decay. The corresponding $t \bar{t}$
branching fractions ({\it B}) are 1.58\%, 3.16\%, and 1.57\% \cite{PDG} for 
the $e^{+}e^{-}$, $e\mu$, and $\mu^{+}\mu^{-}$ channels, respectively.

The D\O\ detector has a silicon microstrip tracker and a central 
fiber tracker located within a 2~T superconducting solenoidal 
magnet~\cite{run2det}. The surrounding liquid-argon/uranium calorimeter 
has a central cryostat covering pseudo-rapidities $|\eta|$ up to 
$1.1$~\cite{pseudorapidity}, and two end cryostats extending coverage 
to $|\eta|\approx 4$~\cite{run1det}. A muon system~\cite{run2muon} 
resides beyond the calorimetry, and consists of a layer of tracking 
detectors and scintillation trigger counters before 1.8~T toroids, 
followed by two similar layers after the toroids. 
Luminosity is measured using plastic scintillator arrays located 
in front of the end cryostats. The trigger and data acquisition systems 
are designed to accommodate the high luminosities of Run II. The data 
used in this analysis were collected by
requiring two leptons ($e$ or $\mu$) in the hardware trigger and one
or two leptons in the software triggers~\cite{run2det}.

To extract the $\ttbar$ signal, we select events with two high-$p_T$
isolated leptons, large $\met$, and at least two jets. We further improve
the signal to background ratio by selecting events with kinematics compatible 
with $\ttbar$ events. To derive the cross section we determine the overall 
efficiency $\epsilon$ (including trigger, geometrical, and event selection 
efficiencies) for $\ttbar$ and the number of expected background events.
We distinguish two categories of backgrounds: ``physics'' and
``instrumental''. Physics backgrounds are processes in which the 
charged leptons arise from electroweak boson decays and the $\met$ 
originates from high $p_T$ neutrinos. This signature arises in 
$Z/\gamma^* \to \tau^+ \tau^-$  where the $\tau$ leptons decay leptonically,
%where $\ell ,\ell' ~=~e~\rm{or}~\mu$ 
and $WW/WZ$ (diboson) production. 
Instrumental backgrounds are defined as events in which (a) a jet or a
lepton within a jet fakes the isolated lepton signature, or (b) the
$\met$ originates from misreconstructed jet or lepton energies or
from noise in the calorimeter.

%%%%%%%%%%%%%%%%%%%%%%%%%

The electrons used in the analysis are defined as clusters of calorimeter 
cells for which (a) the fraction of energy deposited in 
the electromagnetic section of the calorimeter has to be at least 90\% of 
the total cluster energy, (b) the energy is concentrated in a narrow cone 
and isolated from further calorimeter energy, (c) the shape 
of the shower is compatible with that of an electron, (d) the electron 
matches a charged track in the tracking system. In order to further remove 
backgrounds we use (e) a discriminant that selects prompt isolated electrons 
based on the tracking system and calorimeter information~\cite{ljetstopo}. 
Electrons which fulfill criteria (a) to (e) are referred to as ``tight'' 
electrons. For background calculations we introduce ``loose'' electrons for 
which only (a) and (b) are required. The muons considered in the analysis are 
defined as tracks reconstructed in the three layers of the muon system, with 
a matching track in the tracking system. The energy deposited in the calorimeter 
inside a hollow cone around the muon must be less than 12\% of the muon $p_T$. 
To further remove background, the sum of the charged track momenta in a cone 
around the muon track has to be smaller than 12\% of the muon $p_T$. 
Muons that fulfill all these criteria are referred to as ``tight'' muons.
For background  calculations, we introduce ``loose'' muons for which 
the isolation criteria are relaxed.

%%%%%%%%%%%%%55555

Jets are reconstructed with a fixed cone of radius 
$\Delta {\cal R} = 0.5$~\cite{jet} and must be confirmed by the independent 
calorimeter trigger readout. Jet energy calibration is applied to the 
jets \cite{jetscale}. The \met\ is equal in magnitude and opposite in direction to the
vector sum of all significant calorimeter cell transverse energies.  It is
corrected for the transverse momenta of all isolated muons, as well as 
for the corrections to the electron and jet energies.

Event selections for each channel are optimized to minimize the expected
statistical uncertainty on the cross section. We select events with at least two
jets with $p_T^{j} > 20$~GeV and \mbox{$|y|<2.5$}~\cite{pseudorapidity} and 
two leptons with $p_T^{\ell} > 15$~GeV. Muons are accepted in the region 
\mbox{$|\eta| <$ 2.0}, while electrons must be within \mbox{$|\eta| <$ 1.1} 
or \mbox{1.5 $< |\eta| <$ 2.5}. The two leptons are required to be of opposite 
signs in the $e^{+}e^{-}$ and $\mu^{+}\mu^{-}$ channels. 

A cut on \met\ is crucial to reduce the otherwise large $Z/\gamma^*$ background. 
This background is particularly severe in the $e^+ e^-$ and $\mu^+ \mu^-$ channels. 
Due to different resolutions in electron energies and muon momenta, the optimization
leads to different selections in the three channels. 
In the $e\mu$ channel, we require $\met > 25$ GeV and $\Delta\phi(\met,\mu) > 0.25$, 
where $\Delta\phi(\met,\mu)$ is the azimuthal angle between the \met\ and the muon.  
The latter gives additional rejection against $Z/\gamma^* \to \tau\tau$ background
in events with two jets. In the $e^+ e^-$ channel, we veto events with dielectron 
invariant mass  $80 \le M_{ee} \le 100$ GeV and require $\met > 35$ GeV 
($\met > 40$ GeV) for $M_{ee} > 100$ GeV ($M_{ee} < 80$ GeV). In the 
$\mu^+\mu^-$ channel, we accept events with $\met > 35$ GeV. This cut is 
tightened at low and high values of $\Delta\phi(\met,\mu_1)$ where $\mu_1$ denotes
the leading $p_T$ muon.  Events with 
$\Delta\phi(\met,\mu_1) >175 ^{\circ}$ are removed. 

The final selection in the $e\mu$ channel requires 
$H_T^{\ell}=p_T^{\ell_1} + \Sigma(p_T^{j}) > 140$ GeV, where 
$p_T^{\ell_1}$ denotes the $p_T$ of the leading lepton. This cut effectively
rejects the largest backgrounds for this final state which arise from
$Z/\gamma^*\rar\tau^+\tau^-$ and diboson production. The $e^+ e^-$ analysis uses
a cut on sphericity $\mathcal{S} = 3(\epsilon_1 + \epsilon_2)/2 > 0.15$, 
where $\epsilon_1$ and $\epsilon_2$ are the two leading eigenvalues of the 
normalized momentum tensor~\cite{topovar}. This requirement rejects
events in which jets are produced in a planar geometry through gluon
radiation. The final selection applied in the $\mu^+\mu^-$ channel 
further rejects the $Z/\gamma^*\rar \mu^+\mu^-$ background. We compute
for each $\mu^+\mu^-$ event the $\chi^2$ of a fit to the 
$Z\rightarrow \mu^+\mu^-$ hypothesis given the measured
muon momenta and known resolutions. Selecting events with
$\chi^2 > 2$ is more effective than selecting on the dimuon invariant mass 
for this channel.

%%%%%%%%%%%

Signal acceptances and efficiencies are derived from a combination 
of Monte Carlo simulation (MC) and data. Top quark pair production 
is simulated using {\sc alpgen}  \cite{alpgen} with \mtop\ $=175$ GeV. 
{\sc pythia} \cite{pythia} is used for fragmentation and decay. $B$ 
hadron and $\tau$ lepton decays are modeled via {\sc evtgen} \cite{evtgen} 
and {\sc tauola} \cite{tauola}, respectively. A full detector simulation 
using {\sc geant} \cite{geant} is performed. Lepton trigger and 
identification efficiencies as well as lepton momentum resolutions are 
derived from $Z/\gamma^* \rightarrow \ell^+\ell^-$ ($\ell=e,\mu$) data.
These per-lepton normalization factors and momentum smearings are 
applied to MC events to ensure the simulated samples provide an accurate 
description of the data. The jet reconstruction efficiency, jet energy 
resolution and $\met$ resolution in the MC are adjusted to their measured 
values in data.

%%%%%%%%%%%%%%

To calculate the expected number of events from physics backgrounds, 
we use $Z/\gamma^* \to \tau^+ \tau^-$ and diboson MC samples generated 
with {\sc pythia} and {\sc alpgen}, respectively. The 
$Z/\gamma^* \to \tau^+ \tau^-$ contribution is normalized to 
the cross section measured by \dzero\ ~\cite{ztautau}. 
For the diboson processes, diboson + 2 jets events are generated at
leading order (LO) and are scaled by the ratio of the next-to-leading
order to LO inclusive cross sections derived for diboson inclusive
production~\cite{dibosonxs}.

Instrumental backgrounds are determined from the data. Fake electrons  
can arise from jets comprised essentially of a leading $\pi^0$/$\eta$ 
and an overlapping or conversion-produced track. We estimate this background 
by calculating the fraction $f_e$ of loose electrons which appear as tight 
electrons in a control sample dominated by fake electrons. In the $e^+e^-$ 
channel the control sample consists of events that satisfied the trigger 
and have two loose electrons. In the $e\mu$ channel the events in the 
control sample must satisfy the trigger and have one tight muon and one 
loose electron. Contributions from processes with real electrons 
($W\to e \nu$ and $Z/\gamma^* \to e^+e^-$) are suppressed by requiring $\met < 10$ GeV
in both $e^+e^-$ and $e\mu$ channels and $|M_{ee} - M_Z|~>~15$~GeV in 
the $e^+e^-$ channel only. We also veto events in which both loose electrons have a 
matching track. We observe that $f_e$ measured 
in the $e^+e^-$ and $e\mu$ control samples agree within statistical errors.
The predicted number of events with a fake electron in the
final sample is obtained by multiplying the number of $e^+e^-$ 
($e\mu$) events with one loose electron and one tight electron (muon) 
by $f_e$. 

%%%%%%%%%%%%%%%%%%%%%%%%%%%%%%%%%%%%%%%%%%%

An isolated muon can be mimicked by a muon in a jet when the jet 
is not reconstructed. We measure the fraction $f_{\mu}$ of loose 
muons that satisfy the tight muon criteria in a control sample dominated 
by fake muons. In the $\mu^+\mu^-$ channel the control sample is defined 
as events that have two loose muons. To suppress physics processes with 
real isolated muons the leading $p_T$ muon is required to fail the 
tight muon criteria. This 
cuts efficiently $Z/\gamma^*\rar\mu^+\mu^-$ events but also
$W\rar \mu\nu$ events where a second-leading muon might arise from a muon in a jet.
The number of events with a fake muon contributing to the final sample 
is estimated by counting the number of events with one tight muon and a
loose muon and multiplying it by $f_{\mu}$. In the $e\mu$ channel 
the contribution from events where both leptons are fake leptons is 
already accounted for by using $f_e$. The remaining contribution 
from events with a real electron and a fake muon, is determined by 
combining $f_e$ and a fake rate $f_{\mu}$ obtained on
a control sample that satisfies the $e\mu$ trigger.

%%%%%%%%%%%%%%%%%%%%%%%%%%%%%%%%%%%%%%%%%%%

The processes $Z/\gamma^*\rar \ell^+\ell^-$ ($\ell=e,\mu$), while
lacking high $p_T$ neutrinos, might have a significant amount of measured 
$\met$ due to limited $\met$ resolution. In the $e^+e^-$ channel, this 
background is estimated by measuring a \met\ misreconstruction rate on data
and applying it to the simulation. We observe that the \met\ spectrum
in $e^+e^-$ events with $80 \le M_{ee} \le 100$ GeV agrees well with the 
\met\ spectrum observed in $\gamma+ 2\; \rm{jets}$ candidate events. We 
obtain the \met\ misreconstruction rate in data as the ratio of the number of 
$\gamma+ 2\; \rm{jets}$ events passing the \met\ selection divided by the 
number failing the selection. The \met\ misreconstruction rate is also consistent 
with $Z/\gamma^*\rar e^+e^- +2 \;\rm{jets}$ simulation. This rate is multiplied 
by the number of events that fail the \met\ selections but pass all other
selections. In the $\mu^+\mu^-$ channel, the expected contribution of 
$Z/\gamma^*\rar \mu^+\mu^-$ background in the final sample is 
derived from events simulated with {\sc alpgen}. Good agreement 
is observed between the data and the simulation in the variables 
$\met$ and $\Delta \phi(\met,\mu_1)$.  This allows us to obtain the 
probability for a $Z/\gamma^*\rar \mu^+\mu^-$ event to pass the $\met$
selection from the simulation. The sample is normalized to the number 
of observed $Z/\gamma^*\rar \mu^+\mu^-$ events in the data with
$70 \le M_{\mu\mu} \le 110$ GeV before the $\met$ selection.

\begin{figure*}[tbh]
\center{\includegraphics[width=0.99\textwidth]{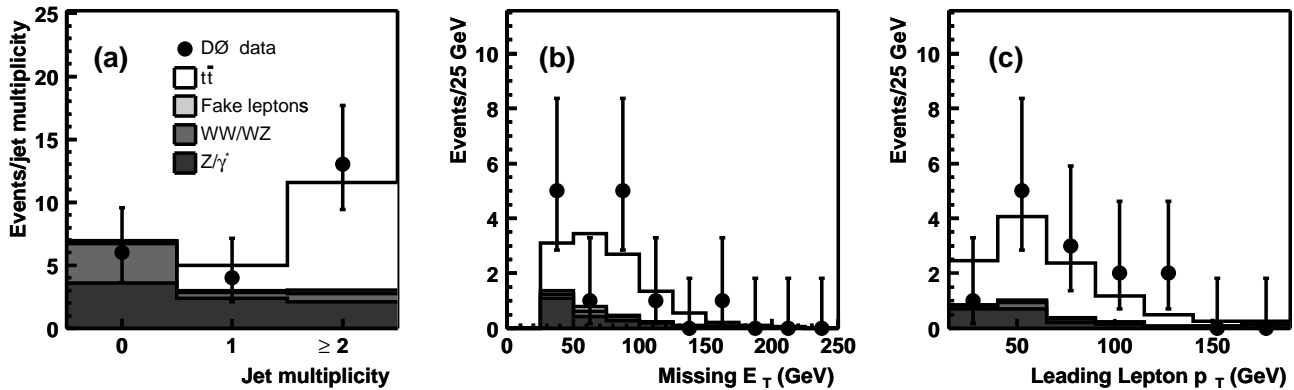}}
\caption{\label{fig:plots} Predicted and observed (a) number of events
with 0, 1 and 2 or more jets with all other selections applied, (b) \met\ 
and (c) leading lepton $p_T$ in dilepton events 
after all selections. The $Z/\gamma^*$ contribution includes $e^{+}e^{-}$, 
$\tau^+\tau^- \to e\mu$, and $\mu^+\mu^-$ final states. The $t\bar{t}$ 
prediction is shown for $\sigma_{t\bar{t}}=7$~pb.}
\end{figure*}

%%%%%%%%%%%%%%%%%%%%%%%%%%%%%%%%%%%%%%%%%%%%%%%%%%%%%%%%%%%%%%%%%%%%%%%%%%%%

The number of observed events and estimated physics and instrumental 
backgrounds in the dilepton + 2 jets sample, the integrated 
luminosities and the $\epsilon\times${\it B} for the \ttbar\
signal are given in Table \ref{tab:evts} for each channel. We observe 5, 8 
and 0 events in the $e^+e^-$, $e\mu$ and $\mu^+\mu^-$ channels, respectively. 
We estimate the probability to observe $\geq 5$, $\geq 8$, and
exactly 0 events in the $e^+e^-$, $e\mu$, and $\mu^+\mu^-$ channels as 22\%,
43\%, and 5\%, respectively, using the measured $\sigma_{t\bar{t}}$ and
taking into account systematic uncertainties. By generating 
pseudo-experiments we estimate that 20\% of the possible outcomes have lower 
likelihoods than that of our observation. The significance of the observed 
$\ttbar$ signal over the background is 3.8 standard deviations.

To compute the cross section, we calculate 
in each channel the probability to observe the number of events seen in 
the data as a function of $\sigma_{t\bar{t}}$ given the number of background 
events and the signal efficiencies. The combined cross section is the value of 
$\sigma_{t\bar{t}}$ that maximizes the product of the likelihoods in the 
three channels. The resulting top quark pair production cross section 
at $\sqrt{s} = 1.96$ TeV in dilepton final states is

\begin{center}
$\sigma_{t\bar{t}}=8.6_{-2.7}^{+3.2}(\rm stat)\pm1.1(\rm syst)\pm0.6(\rm
lumi)$ pb \\
\end{center}
\noindent for $\mtop = 175$ GeV, within errors of the standard model 
theoretical prediction of $6.77\pm0.42$ pb~\cite{cseccalc} and in agreement
with the recent result in Ref.~\cite{cdfllcsec}.
We find $\sigma_{t\bar{t}}$ also consistent with measurements carried out
in different final states~\cite{ljetstopo,topxsother}.
The total systematic uncertainty is obtained by varying the background 
prediction and signal efficiencies within their uncertainties and taking 
into account correlations. The dominant systematic 
uncertainties are given in Table~\ref{tab:syst}. In addition a 6.5\% systematic
uncertainty is assigned to the luminosity measurement~\cite{lumi.}.
The top quark mass affects the signal efficiency, resulting in
a dependence of  $\sigma_{t\bar{t}}$ on $m_{t}$ given by 
$d\sigma_{t\bar{t}}/dm_t =-0.08$~pb/GeV for $m_{t}$ in the range 
160~GeV to 190 GeV.

%%%%%%%%%%%%%%%%%%%%%%%%%%%%%%%%%%%%%%%%%%%%%%%%%%555

Figure~\ref{fig:plots}(a) shows that the observed 
number of events with 0, 1, and 2 or more jets, 
with all other selections applied, is consistent with the prediction 
(assuming $\sigma_{t\bar{t}}=7$ pb). Figure~\ref{fig:plots}(b) 
shows that the observed and predicted \met\ spectra after all selections 
agree well. Other kinematic distributions in dilepton events are also 
well described by the sum of $\ttbar$ signal and background contributions 
at various steps of the event selection. 

The leading lepton $p_T$ spectrum in the $\ttbar$ dilepton final 
states has recently been studied by the CDF Collaboration~\cite{cdflpt} 
and a mild excess has been observed at low transverse momenta. This is 
not confirmed by our data, as shown in Fig.~\ref{fig:plots}(c). To test 
agreement between data and the prediction, we generate pseudo-experiments 
from the predicted leading lepton $p_T$ spectrum and use our measured 
$\sigma_{t\bar{t}}$ to normalize the $\ttbar$ signal. We find that 31\% 
of the pseudo-experiments are less consistent with the parent distribution 
than the data. We conclude that data agree well with the prediction.

In summary, we have measured the top quark pair production cross section 
at $\sqrt{s} = 1.96$ TeV in $e^+e^-$, $e\mu$ and $\mu^+\mu^-$
final states to be
$\sigma_{t\bar{t}}=8.6_{-2.7}^{+3.2}(\rm stat)\pm1.1(\rm syst)\pm0.6(\rm
lumi)$ pb for $\mtop = 175$ GeV, in agreement with the standard model
prediction and with measurements in other final states.

\begin{table}
  \caption{\label{tab:evts} Expected signal (assuming $\mtop = 175$ GeV and 
    $\sigma_{t\bar{t}} =7$ pb) and background event yields for 
    $e^+e^-$, $e\mu$, and $\mu^+\mu^-$ channels. Instrumental 
    backgrounds include \met\ and fake lepton backgrounds. Total uncertainties
    are given.}
  \begin{ruledtabular}
    \begin{tabular}{lccc}
      Channel & $e^+e^-$ & $e\mu$ & $\mu^+\mu^-$ \\
      %\tabularnewline
      \hline
      Integrated luminosity (pb$^{-1}$) & 243 & 228 & 224 \\
      &&& \\
      Physics backgrounds             & $0.3\pm0.1$ &
      $0.7\pm0.2$  & $0.2\pm0.1$ \\
      %\tabularnewline
      Instrumental backgrounds        & $0.7\pm0.1$          &
      $0.2\pm0.1$           & $1.1^{+0.4}_{-0.3}$ \\
      %\tabularnewline
      Total background                & $0.9\pm0.1$          &
      $0.9\pm0.2$  & $1.4\pm0.4$ \\
      %\tabularnewline
      &&& \\
      %\tabularnewline
      $\epsilon \times${\it B} ($10^{-3}$) 
      & $1.1^{+0.1}_{-0.2}$ &
      $3.2^{+0.4}_{-0.3}$  & $1.0\pm0.1$ \\
      %\tabularnewline
      Expected signal                 & $1.9^{+0.2}_{-0.3}$ &
      $5.1^{+0.6}_{-0.5}$  & $1.6\pm0.2$ \\
      %\tabularnewline
      &&& \\
      %\tabularnewline
      Total prediction                & $2.8\pm0.3$ &
      $6.1^{+0.6}_{-0.5}$  & $2.9\pm0.6$ \\
      %\tabularnewline
      \hline
      Observed                        & 5 & 8 & 0 \\
      %\tabularnewline
    \end{tabular}
  \end{ruledtabular}
\end{table}

\begin{table}
\caption{\label{tab:syst} Summary of systematic uncertainties on 
$\sigma_{t\bar{t}}$.}
\begin{ruledtabular}
\begin{tabular}{lc}
Source & {$\Delta\sigma_{t\bar{t}}$ (pb)}\\
\hline
Jet energy calibration  & $+\; 0.8\; -0.7$ \\
Jet identification      & $+\; 0.3\; -0.6$ \\
Muon identification 	& $+\; 0.5\; -0.4$ \\
Electron identification & \multicolumn{1}{c}{$\pm \; 0.3$} \\
Trigger                 & $+\; 0.3\; -0.2$ \\
Other			& $+\; 0.2\; -0.3$ \\
\hline
Total                   & \multicolumn{1}{c}{$\pm \; 1.1$} \\
\end{tabular}
\end{ruledtabular}
\end{table}

% acknowledgement_paragraph_r2.tex                5/17/05
%
We thank the staffs at Fermilab and collaborating institutions, 
and acknowledge support from the 
DOE and NSF (USA);
CEA and CNRS/IN2P3 (France);
FASI, Rosatom and RFBR (Russia);
CAPES, CNPq, FAPERJ, FAPESP and FUNDUNESP (Brazil);
DAE and DST (India);
Colciencias (Colombia);
CONACyT (Mexico);
KRF (Korea);
CONICET and UBACyT (Argentina);
FOM (The Netherlands);
PPARC (United Kingdom);
MSMT (Czech Republic);
CRC Program, CFI, NSERC and WestGrid Project (Canada);
BMBF and DFG (Germany);
SFI (Ireland);
Research Corporation,
Alexander von Humboldt Foundation,
and the Marie Curie Program.
%
   % input acknowledgement


\begin{thebibliography}{99}

% list_of_visitor_addresses_r2.tex                            11/15/04
%
%\bibitem[\dag]{name}
\bibitem[*]{lehner}
Visitor from University of Zurich, Zurich, Switzerland.
%
\vskip 0.25cm


\bibitem{disc}
CDF Collaboration,  F.\ Abe {\it et al.},    Phys. Rev. Lett. {\bf 74}, 2626 (1995);
D\O\ Collaboration, S.\ Abachi {\it et al.}, Phys. Rev. Lett. {\bf 74}, 2632 (1995).

\bibitem{FCNC}
A.\ Bean {\it et al.}, Phys. Rev. D {\bf 35}, 3533 (1987).

\bibitem{T3b}
E.\ Elsen {\it et al.},   Z. Physik C {\bf 46}, 349 (1990); 
H.\ Behrend {\it et al.}, Z. Physik C {\bf 47}, 333 (1990); 
A.\ Shimonaka {\it et al.}, Phys. Lett. B {\bf 268}, 457 (1991).

\bibitem{mtopcombo}
CDF and D\O\ collaborations, TEVEWWG, 
%``Combination of CDF and D\O\ Results
%on the Top-Quark Mass'', 
hep-ex/0404010.

%\bibitem{ewsb} 
%B.\ Dobrescu, Phys. Lett. B {\bf 461}, 99 (1999);
%B.\ Dobrescu, C.\ T.\ Hill, Phys. Rev. Lett. {\bf 81}, 2634 (1998);
%C.\ T.\ Hill, S. Parke, Phys. Rev. D {\bf 49}, 4454 (1994).

\bibitem{charged_higgs}
  J.F. Gunion {\it et al.},
  {\it The Higgs Hunters Guide} (Addison-Wesley, Redwood City, California, 1990), p. 200.

%\bibitem{cseccalc} R.\ Bonciani {\it et al.}, Nucl. Phys. {\bf B529}, 424 (1998);
%	           N.\ Kidonakis and R.~Vogt, Phys. Rev. D {\bf 68}, 114014 (2003);
%                   M.\ Cacciari {\it et al.}, JHEP {\bf 404}, 68 (2004).

\bibitem{S-W} 
A.\ Salam, {\em Elementary Particle Theory} (Stockholm, Almquist and Wiksells, 1967); 
S.\ Weinberg, Phys. Rev. Lett. {\bf 19}, 1264 (1967).

\bibitem{PDG}
S.\ Eidelman {\it et al.}, Phys. Lett. B \textbf{592}, 1 (2004).

\bibitem{run2det}
  D\O\ Collaboration, V.\ Abazov {\it et al.}, {}``The Upgraded D\O\ Detector,''
  in preparation for submission. to Nucl. Instrum. Meth. Phys. Res. A.

\bibitem{pseudorapidity} 
  Rapidity $y$ and pseudo-rapidity $\eta$ are defined as functions of
  the polar angle $\theta$ and parameter $\beta$ as $y(\theta, \beta) \equiv \frac{1}{2} \ln
  {[(1+\beta\cos{\theta})/(1-\beta\cos{\theta})]}$; $\eta(\theta)
  \equiv y(\theta,1)$, where $\beta$ is the ratio of a particle's
  momentum to its energy.

\bibitem{run1det} D\O\ Collaboration, S.\ Abachi {\it et al.},
  Nucl. Instrum. Methods Phys. Res. A {\bf 338}, 185 (1994). 

\bibitem{run2muon} V. Abazov {\it et al.}, 
FERMILAB-PUB-05-034-E (2005)

\bibitem{ljetstopo} D\O\ Collaboration, V.\ Abazov {\it et al.}, hep-ex/0504043.

\bibitem{jet}
Jets are defined using the iterative seed-based cone algorithm with 
$\Delta {\cal R}=\sqrt{(\Delta\phi)^2+(\Delta\eta)^2}=0.5$ (where $\phi$ is 
the azimuthal angle),
including mid-points as described in Sec.~3.5 (p.\ 47) of G.~C.~Blazey {\it
et al.}, in {\it Proceedings of the Workshop: ``QCD and Weak Boson
Physics in Run~II,''} edited by U.~Baur, R.\ K.~Ellis, and
D. Zeppenfeld, FERMILAB-PUB-00-297 (2000).

\bibitem{jetscale} D\O\ Collaboration, B.\ Abbott {\it et al.},
  Nucl. Instrum. Meth. A {\bf 424}, 352 (1999).

\bibitem{topovar} 
%J.D. Bjorken and S.J. Brodsky, Phys. Rev. D {\bf 1}, 1416 (1970);
  V.\ Barger, J.\ Ohnemus, and R.\ J.\ N.~Phillips, Phys. Rev. D {\bf 48}, 3953 (1993).

\bibitem{alpgen} 
  M.\ L.~Mangano {\it et al.}, JHEP {\bf 07},001 (2003).
  
\bibitem{pythia}
  T.\ Sj\"ostrand {\it et al.}, Comput. Phys. Commun. {\bf 135}, 238 (2001).

%\bibitem{evtgen} D.\ Lange {\it et al.}, 
%  The EvtGen Event Generator Package, in Proc. of CHEP (1998).
\bibitem{evtgen} D.\ J.\ Lange, Nucl. Instrum. Methods Phys. Res. A {\bf 462},
152 (2001).

\bibitem{tauola} 
  S.\ Jadach {\it et al.}, Comp. Phys. Commun. \textbf{76}, 361 (1993).

\bibitem{geant}
  R.\ Brun and F.\ Carminati, CERN Program Library Long Writeup W5013, 1993 (unpublished).

\bibitem{ztautau}
  D\O\ Collaboration, V.\ Abazov {\it et al.},
Phys. Rev. D \textbf{71}, 072004 (2005).

\bibitem{dibosonxs} 
  J.\ M.~Campbell and R.\ K.~Ellis, Phys. Rev. D \textbf{60}, 113006 (1999).

\bibitem{cseccalc} R.\ Bonciani {\it et al.}, Nucl. Phys. {\bf B529}, 424 (1998);
	           N.\ Kidonakis and R.~Vogt, Phys. Rev. D {\bf 68}, 114014 (2003);
                   M.\ Cacciari {\it et al.}, JHEP {\bf 404}, 68 (2004).

\bibitem{cdfllcsec} 
  CDF Collaboration, D.\ Acosta {\it et al.}, Phys. Rev. Lett. \textbf{93}, 142001 (2004).

\bibitem{topxsother} 
  CDF Collaboration, D.\ Acosta {\it et al.}, Phys. Rev. D {\bf 71}, 052003 (2005);
  CDF Collaboration, D.\ Acosta {\it et al.}, Phys. Rev. D {\bf 71}, 072005 (2005);
  D\O\ Collaboration, V.\ Abazov {\it et al.}, hep-ex/0504058;
  CDF Collaboration, D.\ Acosta {\it et al.},  hep-ex/0504053.

%D\O\ Collaboration, V.\ Abazov {\it et al.}, submitted to Phys. Rev. Lett., hep-ex/0504043;
%  CDF Collaboration, D.\ Acosta {\it et al.}, submitted to Phys. Rev. D, hep-ex/0504053;
%  CDF Collaboration, D.\ Acosta {\it et al.}, Phys. Rev. D {\bf 71}, 072005 (2005);
%  CDF Collaboration, D.\ Acosta {\it et al.}, Phys. Rev. D {\bf 71}, 052003 (2005).

\bibitem{lumi.}
  T.\ Edwards {\it et al.}, FERMILAB-TM-2278-E (2004).

\bibitem{cdflpt}
  CDF Collaboration, D.\ Acosta {\it et al.}, hep-ex/0412042.
%  ``Search for Anomalous Kinematics in ttbar Dilepton Events at CDF II'', 
%  submitted to Phys. Rev. Lett., hep-ex/0412042.

\end{thebibliography}
\end{document}